\documentclass[12pt]{iopart}
\usepackage[utf8x]{inputenc}
\usepackage{graphicx}
\usepackage{hyperref}
\usepackage{cleveref}
\usepackage{bm}
\usepackage{color}

\usepackage{iopams}
\begin{document}

\title{Rotating Hybrid Stars with Color-Flavor-Locked Quark Matter}

\author{Debashree Sen$^1$ and Gargi Chaudhuri$^{1,2}$}

\address{$^1$Physics Group, Variable Energy Cyclotron Centre, 1/AF Bidhan Nagar, Kolkata 700064, India}
\address{$^2$Homi Bhabha National Institute, Training School Complex, Anushakti Nagar, Mumbai 400085, India}
\ead{debashreesen88@gmail.com, gargi@vecc.gov.in}
\vspace{10pt}
\date{\today}

\begin{abstract}

 In the present work we have achieved phase transition from $\beta$ stable hadronic matter to color-flavor locked (CFL) quark matter with Maxwell construction. The hybrid equation of state (EoS), obtained for different values of bag pressure $B$ and gap parameter $\Delta$, have been used to compute the speed of sound in hybrid star (HS) matter. The structural properties of the HSs in both static and rotating conditions have been calculated in the light of various constraints from different astrophysical and empirical perspectives. The effects of $B$ and $\Delta$ on the EoS and structural properties have been investigated. At a certain density, shortly after phase transition, the HSs become unstable. In static conditions, the mass-radius solutions satisfy the constraints from GW190425, NICER experiment for PSR J0030+0451 and PSR J0740+6620 and from massive pulsars like PSR J0348+0432 and PSR J0740+6620. In rapidly rotating conditions at Kepler frequency, the constraints on maximum mass from the secondary component of GW190814 and that on rotational frequency from fast pulsars like PSR B1937+21 and PSR J1748-2446ad are all satisfied. In slowly rotating conditions, the universality relations in terms of normalized moment of inertia also hold quite well for most of our HS configurations.

\end{abstract}

%
\noindent{\it Keywords}: Hybrid star, Color-Flavor-Locked Quark Phase, Phase transition
%
%
%

\section{Introduction}
\label{Intro}
 
 It has been very challenging till date, from experimental perspectives to examine matter at densities (5 - 10 times nuclear density $\rho_0$) relevant to the core of the neutron stars (NSs). Thus the composition of NS matter (NSM) and its equation of state (EoS) are largely determined by theoretical modeling of NSM, constrained by the results from various astrophysical observations and empirical techniques. The detection of massive pulsars like PSR J0348+0432 \cite{Ant} and PSR J0740+6620 \cite{Fonseca2021}, gravitational wave events GW170817 \cite{GW170817}, GW190425 \cite{GW190425} and GW190814 \cite{GW190814} and the NICER measurements of gravitational mass and radius of PSR J0030+0451 \cite{Riley2019,Miller2019} and PSR J0740+6620 \cite{Miller2021,Riley2021} have constrained the EoS to considerable extent. In rotating conditions, the constraints on rotational frequency is obtained from rapidly rotating pulsars like PSR B1937+21 \cite{Backer} and PSR J1748-2446ad \cite{Hessels}. Moreover, in the slow rotational conditions ($P\leq 10s$) the universal relations in terms of normalized moment of inertia, quadrupole, tidal deformability and compactness also set constraints on the NS/compact star EoS \cite{Yagi,Breu}. The detection of GW190814 \cite{GW190814} has not confirmed whether its secondary component, having mass $2.59^{+0.08}_{-0.09} M_{\odot}$ \cite{GW190814}, is a NS or a black hole. With the use of various models and parameterizations, for different compositions and choice of interactions, several works \cite{Roupas,Blaschke2,Li2021,Ju,Horvath,GW190814_stat} have suggested that the secondary component of GW190814 may be a non-rotating NS/HS. On the other hand, \cite{Zhang,Biswas,GW190814_rot} have suggested that it may be a rapidly rotating pulsar. Both \cite{Zhang,Biswas} predicted its rotational frequency by considering it to be a massive pulsar. 
 
 Theoretical studies have suggested that in dense core of NSs, hadronic matter can undergo phase transition to deconfined or unpaired quark matter (UQM) forming hybrid stars (HSs) \cite{Blaschke2,Glendenning,Zdunik,Sen,Sen6,Bhattacharya,Han,
HS,Liu,Largani}. It is also suggested that at high density, such deconfined quarks can form Cooper pairs near the Fermi surface, the pairing strength being controlled by the gap parameter $\Delta$. Such pairing leads to the formation of color-flavor locked (CFL) quark phase \cite{CFL} and consequent formation of pure CFL stars \cite{Horvath,Drago2004,Linares,CFL_star,CFL_star2,Alford2003} and CFL magetars \cite{Wen,Liang}. Recent works \cite{Roupas,Li2021,Paulucci2017,EnPingZhou,Lourenco} suggested that with proper choice of $\Delta$ and the bag pressure $B$, the CFL stars and their EoS can successfully satisfy the recent constraints from gravitational wave observations and NICER results. Depending on the values of $B$ and $\Delta$, the formation of superconducting quarks is suggested also in HSs \cite{Alford2003,Sharma,Agrawal2009,Agrawal2010,Bonanno, Ayvazyan,Alf-Sed,Curin,BlaschkeCFL}. With the pairing gap $\Delta=$0, the CFL quark matter reduces to UQM.
 
 In the present work, we invoke phase transition of $\beta$ stable hadronic matter to CFL quark matter following Maxwell construction \cite{Sen,Bhattacharya,Han,Agrawal2010,Bonanno,Ayvazyan,Curin}. Hadron-quark phase transition in HSs is achieved mainly with the help of Gibbs and/or Maxwell constructions depending on the value of the surface tension at the transition boundary. Beyond a limiting value ($\geq$ 70 MeVfm$^{-2}$) of the surface tension, the hadron-quark mixed phase obtained with Gibbs construction, becomes mechanically unstable. Under such circumstances, Maxwell construction is a more physically justified and relevant way of achieving phase transition \cite{Maruyama}. However, the value of the surface tension at the hadron-quark interface is still unknown. In the present work we assume that the surface tension at the interface is high enough to ensure that rapid transition occurs at a sharp interface and therefore we employ Maxwell construction to achieve phase transition. For the pure hadronic phase, we adopt the NL3$\omega\rho$6 model that yields a stiff hadronic EoS \cite{Horowitz,Pais}. Other hadronic models yielding stiff EoS may also be employed for the purpose \cite{Ju,Liu,Drago2004,Agrawal2009,Agrawal2010,Curin}. Since the high density environment of NS cores are theoretically favorable for the formation of hyperons, several works \cite{Sharma,Bonanno,Curin} have included them in the hadronic sector. However, their presence is known to soften the EoS thereby reducing the maximum mass of the HSs and the hyperon couplings are not well-determined. Therefore, similar to \cite{Ju,Liu,Agrawal2009,Agrawal2010}, we do not include the hyperons in the hadronic sector of the present work.
 
 For the CFL quark phase, we follow the treatment proposed by \cite{Alford2003} which consider equal density of u, d and s quarks. The formulation of the CFL model in \cite{Alford2003} is a simple one with certain assumptions. It has been made sophisticated over the years in many rigorous works \cite{Buballa} which also brought out effects like the BCS-like energy spectra of quarks and the dependence of CFL phase formation on quark masses. Moreover, in \cite{Alford2003} the simplistic EoS of the CFL phase was formulated on the basis of the ansatz that in the limit of small mass of the s quark, the number of chemical potentials reduces. Later rigorous works \cite{Buballa} obtained further realistic EoS. Also, several works later included the scalar and vector mediators in the quark phase \cite{Linares,Buballa,Steiner,Ferrer,LLopes}. In such works the pairing gaps and the effective quark masses were dynamically generated from the mean field equations \cite{Linares,Steiner,Wu}. Various works also showed the possibility of CFL phase being preceded by the formation of two-flavor paired superconducting quark 2SC phase \cite{Agrawal2010,Bonanno,Ayvazyan,Alf-Sed,Curin,Buballa}. However, in the present work we study direct transition from hadronic phase to CFL phase at high density. In the present work we adopt the formalism depicted in \cite{CFL} as also considered by \cite{CFL_star} and even recently by \cite{Roupas} to describe the possible existence of CFL stars. In this work, we intend to show that by adopting even such a simplistic formalism for the CFL phase \cite{Alford2003}, with proper choice of the gap parameter and the bag constant and a relatively stiff hadronic EoS, the obtained HS configurations still satisfy the present day astrophysical and empirical constraints reasonably well in both static and rotating conditions. In the present work the CFL phase is charge neutral due to equal number density of the individual u, d and s quarks. Hence we do not involve the contributions of the Goldstone bosons and the leptons in order to ensure charge neutrality of the CFL phase. This is because in the present work we have chosen Maxwell construction to obtain hybrid EoS and Maxwell criteria demands local charge neutrality of the individual phases \cite{Glendenning,Sen,Bhattacharya}. We obtain the hybrid EoS for different values of $B$ and $\Delta$. Although \cite{EnPingZhou,Nandi,Nandi2} have put constraint on the value of $B$ with specific models in the light of GW170817 results, the values of $B$ and $\Delta$ are actually still inconclusive. With the obtained hybrid EoS, we have computed the structural properties of the HSs in both static and rotating conditions. The paper is organized as follows. In the next section \ref{Formalism}, we have discussed the main features of the pure CFL quark phase and the mechanism of phase transition with Maxwell construction. We then present our results and relevant discussions in section \ref{Results}. We have summarized and concluded in the final section \ref{Conclusion} of the paper.


\section{Formalism}
\label{Formalism}

 For the pure hadronic phase, we have adopted the NL3$\omega\rho$6 model \cite{Horowitz} in $\beta$ equilibrated condition. The saturation properties of this model are reasonably acceptable as shown in \cite{Fortin,Grill}. This model has been widely explored to understand various compact star properties \cite{Nandi,Fortin2021,Pais}.
  
  Theoretically, at high density relevant to NS cores, there may be formation of the hyperons leading to to softening of the EoS and reduction in maximum mass of the NSs \cite{Glendenning}. However, there is no experimental or observational evidence till date to support the theory of their presence in NS cores. Under such circumstance, the $\beta$-equilibrated matter consisting of the nucleons, electrons and muons forms the most fundamental \cite{Glendenning} and widely considered composition of NSM. Moreover, not all the hyperon couplings in the hadronic sector are well-determined. Thus similar to \cite{Ju,Liu,Agrawal2009,Agrawal2010}, in the present work, we do not consider the formation of hyperons in the hadronic sector.
 
 For the pure CFL quark phase, we consider the formalism depicted in \cite{Alford2003}. The thermodynamic potential is given as
 
\begin{eqnarray}
\hspace*{-1cm} \Omega=\frac{6}{\pi^2} \int_0^{\nu} k^2(k-\mu_q)dk + \frac{3}{\pi^2} \int_0^{\nu} k^2(\sqrt{k^2+m_s^2}-\mu_q)dk - \frac{3\Delta^2\mu_q^2}{\pi^2} + B
\label{omg}
\end{eqnarray}

where, the quark chemical potential is given in terms of baryon chemical potential ($\mu_n$) as $\mu_q=\mu_n/3$ and $B$ denotes the bag pressure. Here the contribution from CFL condensate is introduced in the third term of eq. \ref{omg} in terms of the pairing gap parameter $\Delta$. Thus $\Delta=0$ reduces to UQM. 

 With equal number density of the individual u, d and s quarks, the quarks acquire a common Fermi momentum $\nu$. This also ensures the charge neutrality of the pure CFL quark phase. So we have,
 
\begin{eqnarray}
\rho_u=\rho_d=\rho_s=(\nu^3 + 2\Delta^2\mu_q)/\pi^2
\label{rho}
\end{eqnarray}

with 

\begin{eqnarray}
\nu=2\mu_q - \sqrt{\mu_q^2 + \frac{m_s^2}{3}}
\label{nu}
\end{eqnarray}

where, the mass of s quark is taken to be $m_s=$ 100 MeV.
 
 Unlike \cite{Alford2003,Sharma}, we do not include the effects of Goldstone boson and the leptons in the CFL quark phase in order to ensure local charge neutrality of this phase.

 The energy density is given as
 
\begin{eqnarray}
\varepsilon_{CFL}=\Omega - \mu_q \frac{\partial\Omega}{\partial\mu_q}
\label{e}
\end{eqnarray}

while the pressure as

\begin{eqnarray}
P_{CFL}=-\Omega
\label{p}
\end{eqnarray}

 As mentioned in the Introduction section that the above treatment of the CFL phase \cite{Alford2003} is a simplistic one and based on various assumptions. Over the years several modifications were made to make the model more sophisticated and more realistic \cite{Buballa}. However, in the present work, we adopt the simplistic formalism for the CFL phase \cite{Alford2003} and with proper choice of the gap parameter and the bag constant and a relatively stiff hadronic EoS, we proceed to calculate the properties of HSs in both static and rotating conditions. For the purpose we assume the surface tension at hadron-quark boundary to be sufficiently large and follow Maxwell construction to invoke phase transition. According to Maxwell criteria, phase transition occurs when the baryon chemical potential and pressure of each of the individual charge neutral phases become equal i.e,
 
\begin{eqnarray}
\mu_B^H=\mu_B^{CFL} 
\end{eqnarray}

and

\begin{eqnarray}
P_H=P_{CFL} 
\end{eqnarray} 
 
 We compute the hybrid EoS for different values of the bag pressure and the gap parameter. In order to study their individual effects on the HS properties, we have varied the bag pressure and the gap parameter individually.
 
 With the obtained hybrid EoS, we calculate the speed of sound in HS matter (HSM) as
 
\begin{eqnarray}
C_s^2=\frac{dP}{d\varepsilon}
\end{eqnarray}
 
 We then compute the structural properties of the HSs with the obtained hybrid EoS. The static structural properties like the gravitational mass and radius are calculated by integrating the Tolman-Oppenheimer-Volkoff (TOV) equations \cite{tov}. The rotational properties like the rotational gravitational mass, radius, rotational frequency $\nu$ and the moment of inertia $I$ of the HSs are obtained with the help of rotating neutron star (RNS) code \cite{Stergioulas}, which is based on the Komatsu-Eriguchi-Hachisu (KEH) method \cite{Komatsu}. The stability of the star is decided by the limiting frequency of rotation known as the Kepler frequency $\nu_K$, which signifies the balance between the centrifugal force and gravity. A star rotating with frequency beyond $\nu_K$ looses mass (mass-shedding limit) from its equatorial region and eventually becomes unstable \cite{Hartle68}. The degree of deformation and stability of the star is decided by its rotational velocity/frequency. 

\section{Result and Discussions}
\label{Results}
 
 The EoS of the pure hadronic matter, with the chosen NL3$\omega\rho$6 model, is adopted from \cite{Fortin}. For the quark phase, we choose two values of bag pressure $B^{1/4}$ as 185 and 200 MeV. For each of these values of $B$, we consider three values of the gap parameter $\Delta$ as 0 (UQM), 35 and 50 MeV. With these chosen values of $B$ and $\Delta$, we obtain the EoS of CFL quark phase using eqs. \ref{e} and \ref{p}. 
 
\begin{figure}[!ht]
\centering
\includegraphics[width=0.5\textwidth]{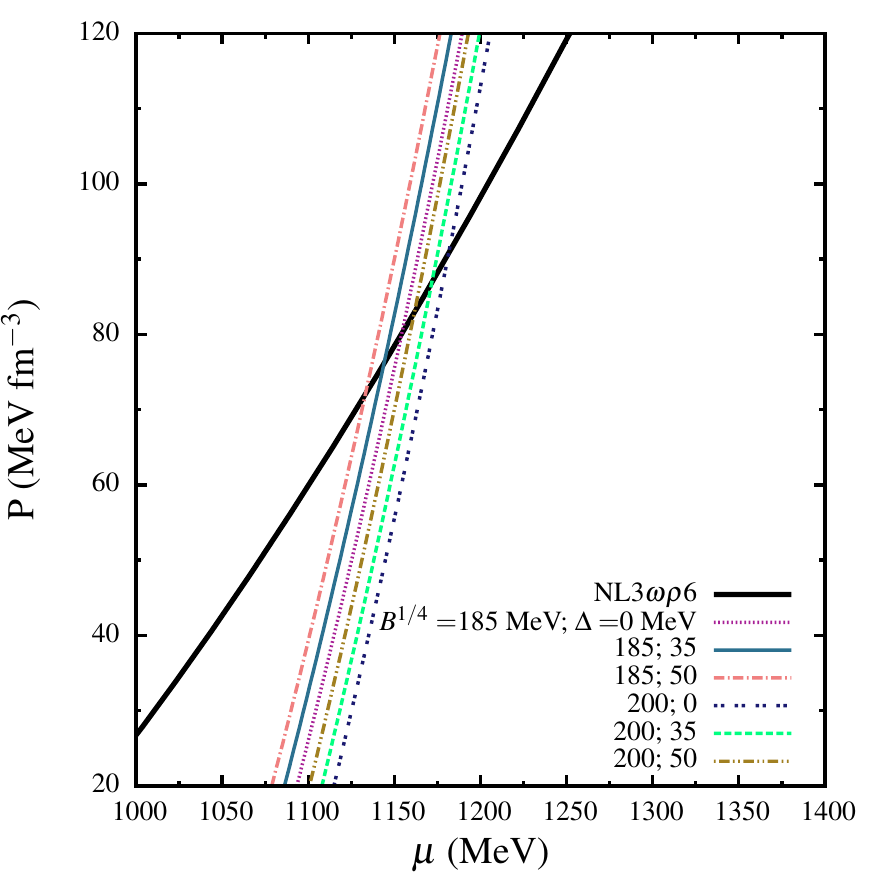}
\caption{Hadronic matter-CFL quark matter crossover for different values of bag constant $B$ and gap parameter $\Delta$.}
\label{cross}
\end{figure}

 In figure \ref{cross}, we compare the pressure as a function of the baryon chemical potential of the individual phases in order to obtain the hadron-quark transition or the crossover points. We tabulate below in table \ref{table_trans} the quantities at transition viz. the transition chemical potential $\mu_t$, transition pressure $P_t$, corresponding transition hadronic density $\rho_t^H$, transition CFL density $\rho_t^{CFL}$, transition hadronic energy density $\varepsilon_t^H$ and the transition CFL energy density $\varepsilon_t^{CFL}$ for the chosen combination of $B$ and $\Delta$.
 
\begin{table}[ht!]
\begin{center}
\caption{Hadron-quark transition properties for the chosen values of bag constant $B$ and gap parameter $\Delta$.}
\setlength{\tabcolsep}{5.0pt}
\begin{center}
\begin{tabular}{cccccccc}
\hline
\hline
\multicolumn{1}{c}{$B^{1/4}$} &
\multicolumn{1}{c}{$\Delta$} &
\multicolumn{1}{c}{$\mu_t$} &
\multicolumn{1}{c}{$P_t$} &
\multicolumn{1}{c}{$\rho_t^H/\rho_0$} &
\multicolumn{1}{c}{$\rho_t^{CFL}/\rho_0$} &
\multicolumn{1}{c}{$\varepsilon_t^H$} &
\multicolumn{1}{c}{$\varepsilon_t^{CFL}$} \\
\multicolumn{1}{c}{(MeV)} &
\multicolumn{1}{c}{(MeV)} &
\multicolumn{1}{c}{(MeV)} & 
\multicolumn{1}{c}{(MeV fm$^{-3}$)} &
\multicolumn{1}{c}{} &
\multicolumn{1}{c}{} &
\multicolumn{1}{c}{(MeV fm$^{-3}$)} &
\multicolumn{1}{c}{(MeV fm$^{-3}$)} \\
\hline
185  &0   &1154.69  &82.14  &2.34  &6.90  &371.0  &1026.94 \\
     &35  &1144.08  &76.52  &2.17  &6.15  &364.0  &1007.93 \\
     &50  &1132.89  &73.40  &2.07  &6.10  &357.0  &993.86 \\

\hline
 
200   &0   &1181.04  &91.16  &2.70  &7.20  &388.0  &1085.67 \\
      &35  &1171.88  &87.14  &2.50  &7.15  &383.0  &1070.25 \\
      &50  &1161.92  &83.49  &2.39  &7.10  &376.0  &1055.36 \\
\hline
\hline
\end{tabular}
\end{center}
\protect\label{table_trans}
\end{center}
\end{table}

 For any particular value of $\Delta$, the crossover shifts to higher values for higher value of $B$ while for a particular value of $B$, higher values of $\Delta$ result in lowering of the values of transition quantities. The difference in values and $\rho_t^{CFL}$ ($\varepsilon_t^{CFL}$) and $\rho_t^H$ ($\varepsilon_t^H$) predicts a large region of phase transition and jump in density.

\begin{figure}[!ht]
\centering
\includegraphics[width=0.5\textwidth]{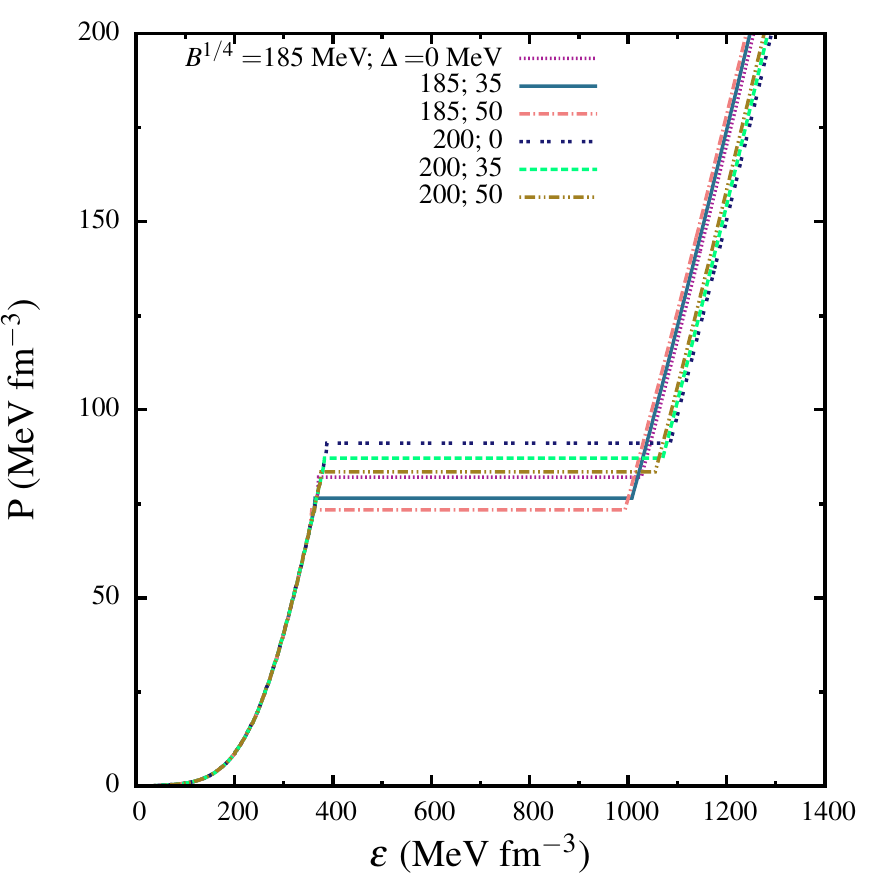}\hfill
\includegraphics[width=0.5\textwidth]{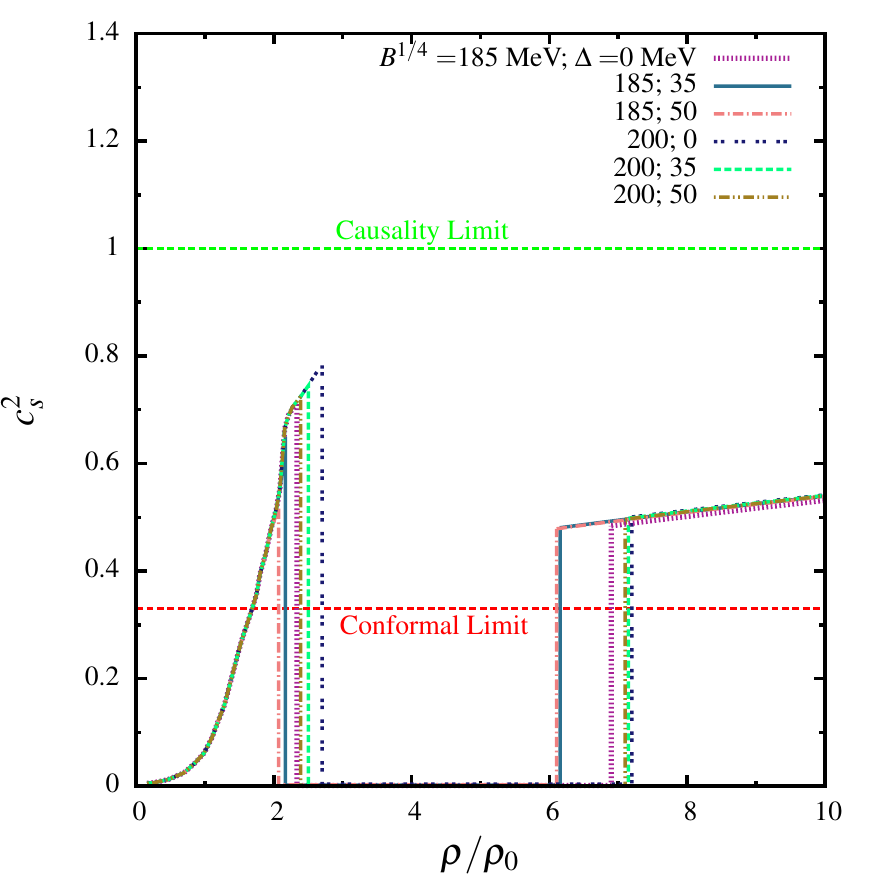}
\caption{Left: Equation of State of hybrid star matter for different values of bag constant $B$ and gap parameter $\Delta$. Right: Speed of sound for the same.}
\label{eos_cs}
\end{figure}

 We then proceed to compute the hybrid EoS for the chosen combination of $B$ and $\Delta$ using Maxwell construction. We display our corresponding results in the left panel of figure \ref{eos_cs} while the right panel of the same figure shows the corresponding variation of speed of sound in the HSM. For all the chosen values of ($B,\Delta$) there is a large jump in energy density indicating the phase transition region (left panel of figure \ref{eos_cs}). As expected, the lower value of $B$ results in early transition compared to the higher value of $B$ while the scenario is just the opposite in case of $\Delta$. As the chosen values of $B$ are quite high, we obtain large density jumps. Since in general, the formation of pure CFL phase occurs at high density, it is justified that there should be adequate difference in density between the existence of pure hadronic and pure CLF quark phases.
 
 As the chosen hadronic model NL3$\omega\rho$6 is very stiff, we find the speed of sound to be quite high in the pure hadronic phase and the peak is noticed just before transition. $C_s^2$ peaks at higher values (average 0.75) for a higher value $B$ ($B^{1/4}=$ 200 MeV) while the maximum value of $C_s^2$ is 0.62 on average for $B^{1/4}=$ 185 MeV. For a particular value of $B$, the value of $C_s$ reduces with higher values of $\Delta$. With hadron-quark phase transition, the speed of sound reduces drastically to $\sim$zero and continues to be so till the onset of the pure CFL quark phase, beyond which the value of $C_s^2$ again rises drastically and then increases in the pure CFL quark phase. This nature of the speed of sound in HSM is consistent with the findings \cite{Zhang,Zdunik,cs,cs2}. Unlike that in the pure CFL phase, the increase in $C_s^2$ much rapid in the pure hadronic phase. In the pure CFL phase the maximum value of $C_s^2$ is 0.53 while in the pure hadronic phase the maximum value is 0.78. However, in both the pure phases and for all the chosen values of $B$ and $\Delta$, the peaks lie above the conformal limit $C_s\leq1/3$ but below the causality limit $C_s\leq1$. In both the pure phases, the speed of sound is slightly higher for higher value of $B$ at any particular value of $\Delta$ while for a fixed value of $B$, it increases slightly with increasing values of $\Delta$. Thus from the present results we notice a highly uphill-downhill behavior of the speed of sound in the HSM. Such behavior is also seen in \cite{Sen6,cs2}.

 With the obtained hybrid EoS, the structural properties are obtained in static conditions using the TOV equations \cite{tov}.

\begin{figure}[!ht]
\centering
\includegraphics[width=0.5\textwidth]{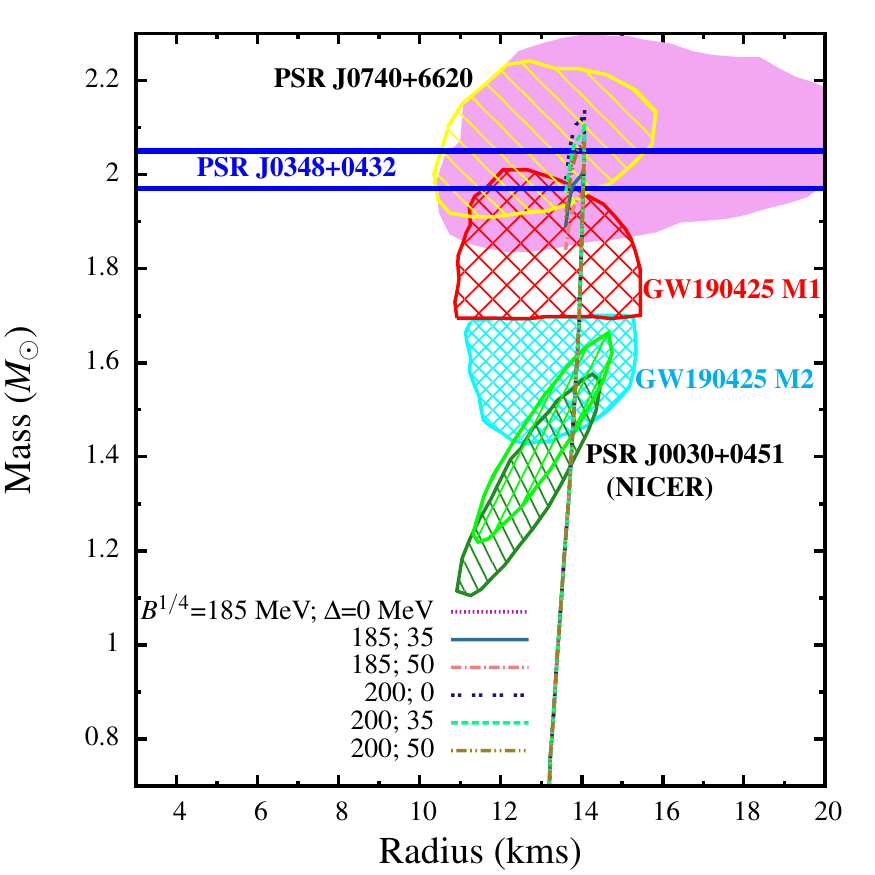}\hfill
\includegraphics[width=0.5\textwidth]{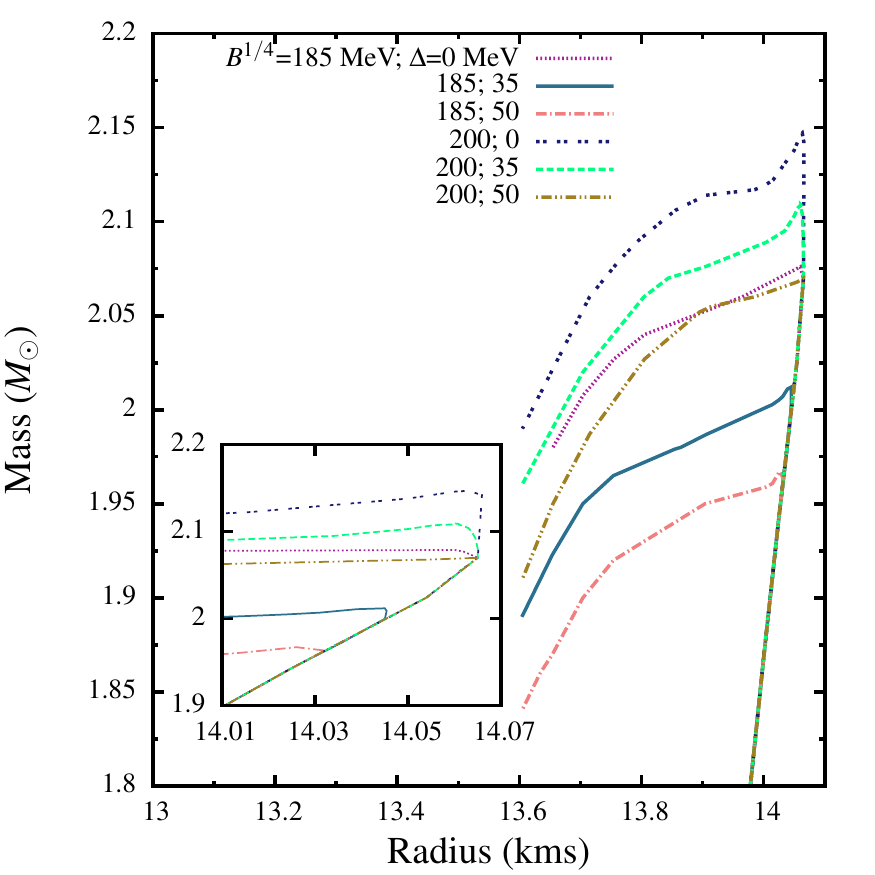}
\caption{Left: Mass-radius relationship of static hybrid star for different values of bag constant $B$ and gap parameter $\Delta$. Observational limits imposed from high mass pulsars like PSR J0348+0432 ($M = 2.01 \pm 0.04 M_{\odot}$) \cite{Ant} and PSR J0740+6620 ($M = 2.08 \pm 0.07 M_{\odot}$ \cite{Fonseca2021} and $R = 13.7^{+2.6}_{-1.5}$ km \cite{Miller2021} or $R = 12.39^{+1.30}_{-0.98}$ km \cite{Riley2021})) are also indicated. The constraints on $M-R$ plane prescribed from GW190425 \cite{GW190425} and NICER experiment for PSR J0030+0451 \cite{Riley2019,Miller2019} are also compared. Right: Mass-radius relationship of static hybrid star for different values of bag constant $B$ and gap parameter $\Delta$ (zoomed portion of transition). The inset highlights the small window within which the different configurations remain stable.}
\label{static}
\end{figure}

 In figure \ref{static}, we show the variation of gravitational mass $M$ with radius $R$ in static conditions. From the left panel, we see that the $M-R$ solutions for all the chosen combinations of ($B^{1/4}$, $\Delta$) except (185, 50) satisfy the constraints on the maximum gravitational mass $M_{max}$ from PSR J0348+0432 \cite{Ant} and PSR J0740+6620 \cite{Fonseca2021}. In the right panel of figure \ref{static}, we have also shown the zoomed portion of $M-R$ plane that is affected by phase transition. The maximum gravitational mass ($M_{max}=$ 2.14 $M_{\odot}$) with corresponding radius ($R_{max}=$ 14.06 km) are obtained with the combination ($B^{1/4}$, $\Delta$)=(200, 0) - a HS configuration with UQM. Among the HS configurations with CFL quark matter, the maximum gravitational mass (2.11 $M_{\odot}$) with corresponding radius 14.06 km is achieved with $B^{1/4}=$ 200 MeV and $\Delta=$ 35 MeV. The lowest value of maximum gravitational mass is 1.96 $M_{\odot}$ with corresponding radius 14.03 km for $B^{1/4}=$ 185 MeV and $\Delta=$ 50 MeV. Overall, for fixed $\Delta$, higher bag pressure leads to comparative delayed transitions (left panel of figure \ref{eos_cs}) and hence more massive HS configurations with slightly greater radius. With fixed value of $B$, the scenario is just the opposite for increasing values of $\Delta$. Although there is substantial variation in $M_{max}$ for the different HS configurations, but the variation in $R_{max}$ is quite less. In the right panel, we can see that soon after phase transition, the $M-R$ configurations become unstable after a certain point i.e, the HSs remain stable within a very small window of the radius. The unstable region corresponds to the region where $dM/d\varepsilon_c <0$, where $\varepsilon_c$ is the central energy density of the star. The inset shows that the only configuration which is completely unstable is for ($B^{1/4}$, $\Delta$)=(200, 50). All the other configurations show that $M_{max}$ is slightly higher than the individual transition points. This rapid unstable nature soon after phase transition may be attributed to rapid and drastic phase transition and considerable high density jump, guided by Maxwell construction. This is also observed in \cite{Curin} where phase transition from hadronic phase to CFL quark phase is studied with the extended FCM model using Maxwell construction. With smooth and slow transition obeying Gibbs construction in place of Maxwell construction, the unstable branch would have not been observed.
 
 From the left panel of figure \ref{static}, it is also clear that the constraints on the $M-R$ plane from GW190425 \cite{GW190425} and NICER experiment for PSR J0030+0451 \cite{Riley2019,Miller2019} are satisfied by all the HS configurations. Our radii estimates $R_{max}$ corresponding to $M_{max}$ for all the chosen combinations of ($B^{1/4}$, $\Delta$) satisfy the recent constraints of NICER experiment for PSR J0740+6620 \cite{Riley2021,Miller2021}. As phase transition occurs at quite high density (and hence high gravitational mass), the value of $R_{1.4}(=$13.75 km) is unaffected by phase transition and hence they remain same for all the chosen combinations of $B$ and $\Delta$ and also same as that obtained with the pure hadronic matter for the chosen NL3$\omega\rho$6 model. 
 

 We next present our results under rotational conditions. We first focus on the structural properties of HSs rotating rapidly at Kepler frequency $\nu$. 

\begin{figure}[!ht]
\centering
\includegraphics[width=0.5\textwidth]{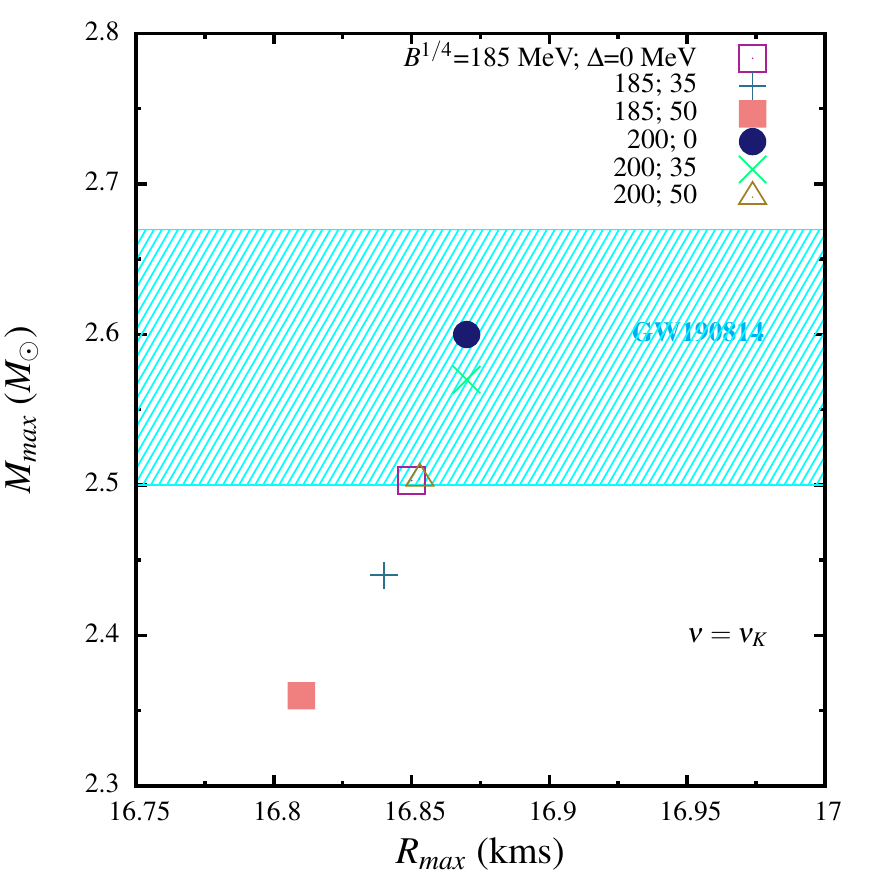}\hfill
\includegraphics[width=0.5\textwidth]{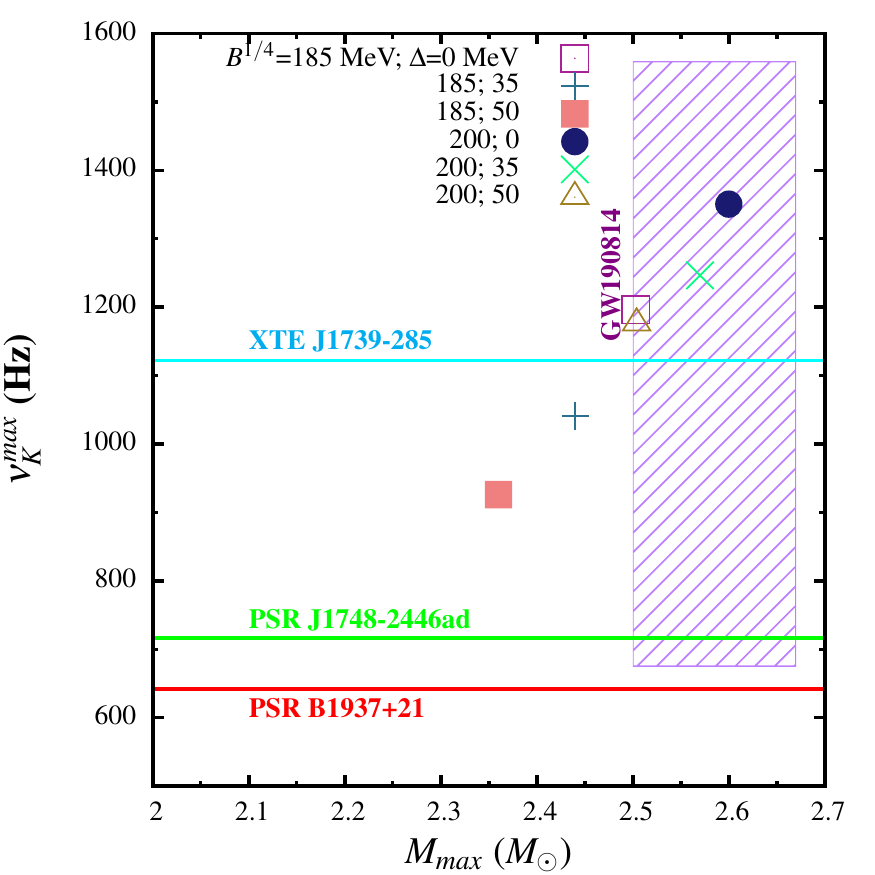}
\caption{Left: Maximum mass vs corresponding radius of hybrid star for different values of bag constant $B$ and gap parameter $\Delta$ rotating at Kepler frequency. Mass of secondary component of GW190814 ($M = 2.59^{+0.08}_{-0.09} M_{\odot}$ \cite{GW190814} - cyan shaded region) is also compared. Right: Maximum mass vs maximum rotational frequency for the same. The frequencies from fast rotating pulsars such as PSR B1937+21 ($\nu = 633$ Hz) \cite{Backer} and PSR J1748-2446ad ($\nu = 716$ Hz) \cite{Hessels} and XTE J1739-285 ($\nu=1122$ Hz) \cite{Kaaret} are also indicated. Result of \cite{Biswas} ($\nu = 1170^{+389}_{-495}$ - purple shaded region) for the secondary component of GW190814 is also compared.}
\label{mrK_mnuK}
\end{figure}

 The variation of maximum gravitational mass $M_{max}$ with corresponding radius $R_{max}$ (left panel) and that of maximum rotational frequency $\nu_K^{max}$ with maximum gravitational mass (right panel) are shown in figure \ref{mrK_mnuK}. As expected, the maximum gravitational mass and corresponding radius is more in case of rotating conditions compared to that of the static case as the centrifugal force comes into play in the former case \cite{Hartle68}. We find that the constraint on maximum mass of the secondary component of GW190814 \cite{GW190814} is satisfied by the HS configurations with both UQM and CFL quark matter for the higher value of $B$. With the lower value of $B$, this constraint is satisfied only by the HS configuration achieved with UQM. 
 
 According to the right panel of figure \ref{mrK_mnuK}, the same conclusion can be drawn in case of satisfying the maximum rotational frequency constraint of the secondary component of GW190814 predicted by \cite{Biswas} assuming it to be a rapidly rotating pulsar. The constraints from fast rotating pulsars like PSR B1937+21 \cite{Backer} and PSR J1748-2446ad \cite{Hessels} are satisfied by all the HS configurations while that from XTE J1739-285 \cite{Kaaret} is satisfied by all the chosen HS configurations except for those with CFL quark matter for $B^{1/4}=$185 MeV.


      
 We next calculate the structural properties for slow rotation at frequency $\nu=$500 Hz.

\begin{figure}[!ht]
\centering
\includegraphics[width=0.5\textwidth]{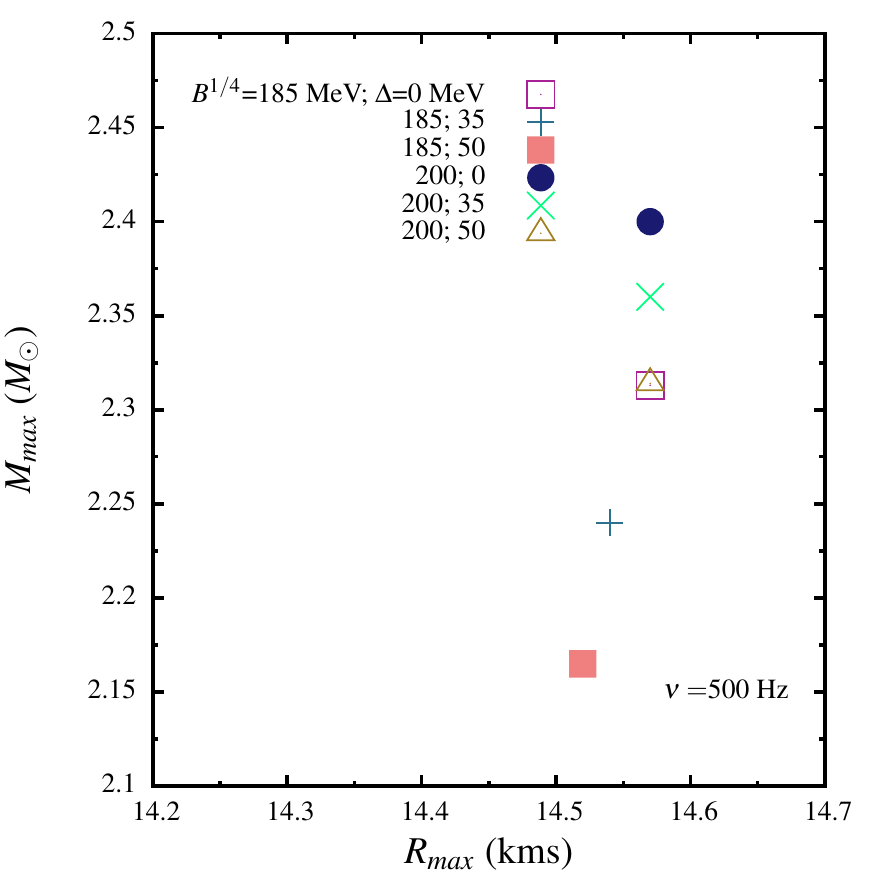}
\caption{Maximum mass vs corresponding radius of hybrid star for different values of bag constant $B$ and gap parameter $\Delta$ rotating at frequency $\nu=$500 Hz.}
\label{mr500}
\end{figure}

 In figure \ref{mr500}, we present the variation in maximum mass with corresponding radius of HSs rotating slowly at frequency $\nu=$500 Hz. Both maximum mass and corresponding radius is less for a lower rotational frequency compared to the case of a higher one.

\begin{figure}[!ht]
\centering
\includegraphics[width=0.5\textwidth]{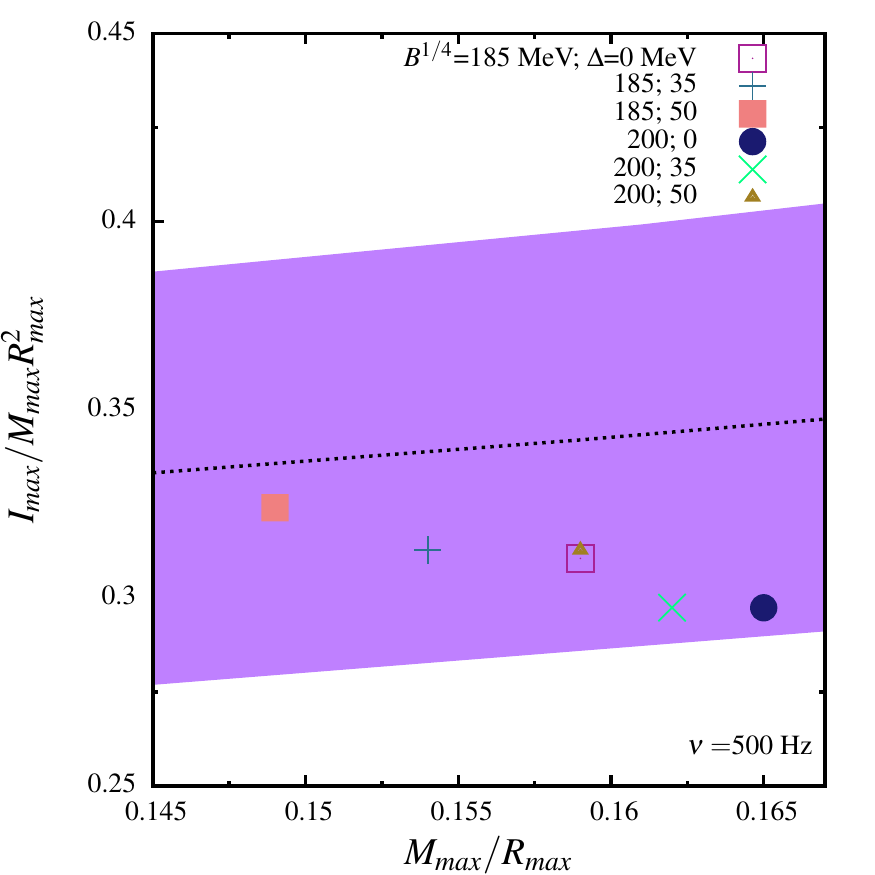}\hfill
\includegraphics[width=0.5\textwidth]{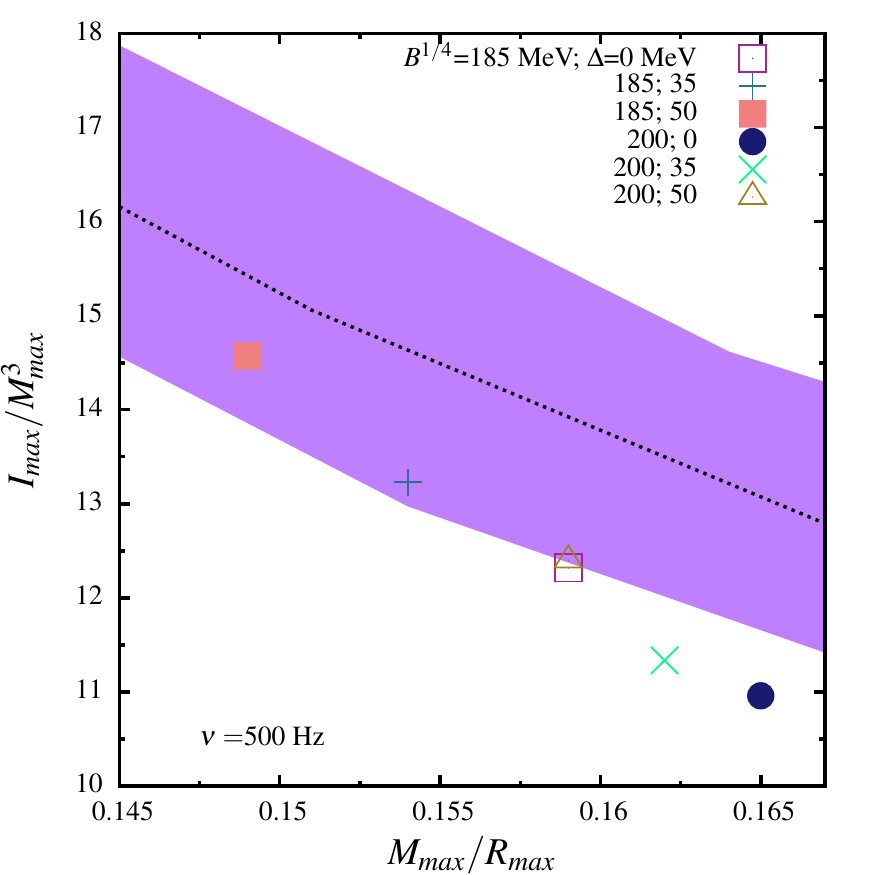}
\caption{Left: Normalized maximum moment of inertia ($I_{max}/M_{max}R_{max}^2$) versus maximum compactness factor ($M_{max}/R_{max}$) of hybrid star for different values of bag constant $B$ and gap parameter $\Delta$ rotating at frequency $\nu=$500 Hz. Right: Normalized maximum moment of inertia ($I_{max}/M_{max}^3$) versus maximum compactness factor ($M_{max}/R_{max}$) for the same. The fitted value of normalized $I$ from various theoretical models for slow rotation (black dashed line) \cite{Lattimer} is shown along with the uncertainty region (shaded region) \cite{Breu}}
\label{CI_CI2}
\end{figure}

 For rotational frequency $\nu=$500 Hz, we compute the moment of inertia of the HSs. In figure \ref{CI_CI2} we 
show the variation of normalized maximum moment of inertia $I_{max}/M_{max}R_{max}^2$ (left panel) and $I_{max}/M_{max}^3$ (right panel) with respect to maximum compactness factor $M_{max}/R_{max}$. Comparing with the constraints of universal relations prescribed by \cite{Breu}, we find that all the HS configurations satisfy this constraint in terms of $I/MR^2$. The constraint for $I/M^3$ is fulfilled by all the HS configurations for $B^{1/4}=$ 185 MeV while for $B^{1/4}=$ 200 MeV the constraint is barely satisfied by the HS configuration obtained only with the maximum value of $\Delta$. We note that the two most massive configurations, obtained with ($B^{1/4}$, $\Delta$)=(200, 0) and (200, 35) do not satisfy the universality relation for $I/M^3$.

 
\begin{figure}[!ht]
\centering
\includegraphics[width=0.5\textwidth]{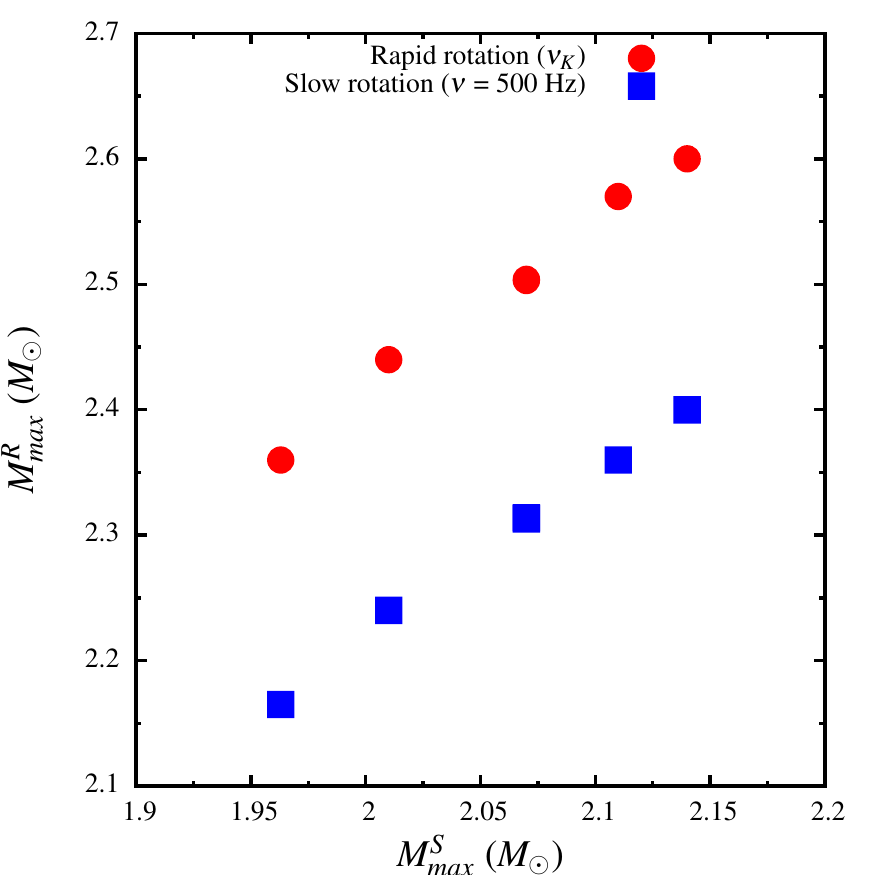}
\caption{Maximum mass in static condition vs maximum mass in rotational condition of hybrid star for different values of bag constant $B$ and gap parameter $\Delta$ for rapid ($\nu=\nu_K$) and slow rotation ($\nu=$500 Hz).}
\label{mSR}
\end{figure}

 We finally compare the change in maximum mass of HSs from static to slow ($\nu=$ 50 Hz) and rapidly ($\nu_K$) rotating conditions in figure \ref{mSR}. Here $M_{max}^S$ and $M_{max}^R$ denote maximum mass at static and rotational conditions, respectively. The average change in maximum mass from static to slow rotation is about 11.5\% while for rapid rotation it is about 21.1\%. From static to slow rotation, the average change in $R_{max}$ is is 3.5\% while for rapid rotation it is 19.8\%. The change in maximum mass from static to rotational condition is dependent on the particular EoS and the rotational frequency considered. It is known that fast rotating NSs can sustain more massive configurations since in the rotating case the centrifugal force, that play important role in determining both mass and radius of NSs, also increases with frequency \cite{Hartle68}. Also, for a particular frequency, the change is different for different EoS. Therefore for certain models with different composition \cite{Mellinger}, the change in maximum mass from static to rotational condition is smaller compared to the results of the present work. Our results in terms of the change in maximum mass from static to rapid rotational condition are quite consistent with that obtained in \cite{Largani} for various models at $T=0$ for both hadronic and HSM. In the present work the average value of $M_{max}^R$/$M_{max}^S$ is 1.21 for rapid rotation (Keplerian limit) which is quite consistent with that obtained in \cite{Largani}.


 

 
\section{Summary and Conclusion}
\label{Conclusion}

 We achieved HS configurations considering Maxwell construction for phase transition from $\beta$ equilibrated hadronic matter (described by NL3$\omega\rho$6 model) to CFL quark matter. Sharp phase transitions with large density jumps (average 4.4$\rho_0$) are seen for chosen the chosen values of bag pressure and gap parameter. In such HSM, just before transition, the speed of sound peaks high in between the conformal and causality limits. Consequently, we studied the structural properties of the HSs in both static and rotating conditions. In static conditions the $M-R$ solutions of the HSs satisfy the constraints from GW190425 and NICER experiments for PSR J0030+0451 and PSR J0740+6620. It is also seen that at particular density (radius) soon after phase transition, the $M-R$ solutions become unstable. For a fixed value of $B$, both the maximum mass and the corresponding radius decrease with increasing values of $\Delta$ while the opposite trend in variation of $M_{max}$ and $R_{max}$ is noticed with respect to $B$ for any particular value of $\Delta$. This also ensures more massive HS configurations with UQM compared to that with CFL quark matter. However, unlike $M_{max}$, the change in $R_{max}$ do not show a substantial variation for different HSs configurations.
 
 We extended the work to compute the structural properties at rotational conditions. Both the cases of slow and rapid rotations are studied. Rotating at Kepler frequency, all the configurations satisfy the constraints on maximum rotational frequency obtained from PSR B1937+21, PSR J1748-2446ad and XTE J1739-285 while most of the HS configurations satisfy the maximum mass constraint obtained from the secondary component of GW190814 and its frequency range predicted by \cite{Biswas}. At slow rotation limit, all the HS configurations satisfy the universality relation predicted in terms of normalized moment of inertia $I/MR^2$ while most configurations satisfy that in terms of $I/M^3$. We also studied the change in maximum mass of the HSs from slow to rapid rotation with respect to static case. We found that substantial change in both $M_{max}$ and $R_{max}$ occurs as the conditions change from static to slow and then rapid rotation. Thus the present work shows that by adopting even a simplistic formalism for the CFL phase \cite{Alford2003}, the obtained HS configurations can still satisfy the present day astrophysical and empirical constraints reasonably well in both static and rotating conditions. For the purpose the gap parameter and the bag constant should be chosen accordingly along with a relatively stiff hadronic EoS.


\ack

The authors thank Dr. Naosad Alam for providing the hadronic EoS.


\end{document}